\documentclass[12pt]{article}
\usepackage{mathtools}

\renewcommand{\d}{{\rm d}}
\renewcommand{\d}{{\rm d}}
\newcommand{\bT}{\overline{T}}
\newcommand{\bx}{\mbox{\boldmath$x$}}
\newcommand{\by}{\mbox{\boldmath$y$}}
\newcommand{\bn}{\mbox{\boldmath$n$}}
\newcommand{\K}{{\cal K}}
\newcommand{\gsimeqscr}{\begin{smallmatrix} > \\ \sim \end{smallmatrix}}
\newcommand{\vf}{\mbox{\boldmath$f$}}

\newcommand{\bDelta}{\overline{\Delta}}
\newcommand{\bp}{\mbox{\boldmath$p$}}
\newcommand{\tbn}{\mbox{\boldmath$\tilde{n}$}}
\newcommand{\tlr}{\tilde{r}}
\newcommand{\ttau}{{\textstyle \tau}}
\newcommand{\trone}{{\textstyle r_1}}
\newcommand{\rl}{\ell}
\begin{document}

\title{On the discrete version of the black hole solution
}

\author{V.M. Khatsymovsky \\
 {\em Budker Institute of Nuclear Physics} \\ {\em of Siberian Branch Russian Academy of Sciences} \\ {\em
 Novosibirsk,
 630090,
 Russia}
\\ {\em E-mail address: khatsym@gmail.com}}
\date{}
\maketitle
\begin{abstract}
A Schwarzschild type solution in Regge calculus is considered. Earlier, we considered a mechanism of loose fixing of edge lengths due to the functional integral measure arising from integration over connection in the functional integral for the connection representation of the Regge action. The length scale depends on a free dimensionless parameter that determines the final functional measure. For this parameter and the length scale large in Planck units, the resulting effective action is close to the Regge action.

Earlier, we considered the Regge action in terms of affine connection matrices as functions of the metric inside the 4-simplices and found that it is a difference form of the Hilbert-Einstein action in the leading order over metric variations between the 4-simplices.

Now we take the (continuum) Schwarzschild problem in the form where spherical symmetry is not set a priori and arises just in the solution, take the difference form of the corresponding equations and get the metric (in fact, in the Lemaitre or Painlev\'{e}-Gullstrand like frame), which is nonsingular at the origin, just as the Newtonian gravitational potential, obeying the difference Poisson equation with a point source, is cut off at the elementary length and is finite at the source.
\end{abstract}

\section{Introduction}

Regge calculus \cite{Regge} is known as a coordinateless discretization of general relativity (GR). In fact, it is the same GR, but on a certain subclass of the Riemannian manifolds, piecewise flat ones, able in a certain sense to approximate any Riemannian space-time with arbitrarily high accuracy \cite{Fein,CMS}. Regge calculus was used in classical numerical applications and in constructing quantum models \cite{WilTuc,RegWil}. In quantum gravity, there are applications related to regularization in approaches with functional integrals, both in those that are closer to standard lattice field theory \cite{Ham}, and in the Causal Dynamical Triangulations approach \cite{cdt}. There is also connection with the spin-foam models \cite{Per} and Loop Quantum Gravity \cite{Dup}.

The Schwarzschild and Reissner-Nordstrøm geometries within Regge calculus were considered in Ref \cite{Wong}. A certain fixed icosahedral decomposition of three-dimensional space into tetrahedra was used there. The task was to provide the best possible approximation to the continuum case. Although effective lattice methods alternative to Regge calculus are also proposed for the numerical study of such systems \cite{Bre2}. Also cosmological models were analysed numerically with the help of Regge calculus [\cite{WilCol} -- \cite{WilLiu}]. The emergence of cosmological models in the Causal Dynamical Triangulations approach was considered in Ref \cite{GlaLol}.

A resolution of the Schwarzschild black hole singularity in Loop Quantum Gravity was given in \cite{Ash1,Ash2}. Although it is a continuum theory, such a resolution eventually occurs due to the {\it discreteness} of the spectra of geometric quantities like area and the existence of their minimal quantum of the order of the Plank scale.

We would like to analyze a possible black hole solution when, due to specific properties of the discrete path integral measure, the background elementary lengths are loosely fixed dynamically at a certain microscopic scale (defined by the Planck scale and some else parameter characterizing a certain freedom in constructing the functional measure), and we should not pass to the continuum limit.

We discuss dynamically fixing background elementary lengths in \cite{our1}. In that paper, we consider a perturbative expansion in Regge gravity and a (loose) fixation of the background configuration for it, in particular, the background length scale, using the functional integral approach. The functional integral measure can be fixed using the canonical Hamiltonian formalism (this implies an intermediate consideration of the simplicial manifold whose edges are arbitrarily strongly shrunk along any given coordinate, which plays the role of a "continuous time" necessary in such a formalism; the requirement for the full discrete measure is that it should pass to the canonical measure whatever coordinate is taken as a time and such a continuous time limit is performed). Initially, we consider an extended set of variables with an orthogonal connection, which is an independent variable in addition to the tetrad type variables (the edge vectors or area tensors). Such a connection and curvature in the Regge calculus (finite rotation matrices) were defined by Fr\"{o}hlich \cite{Fro}.

In the continuum GR, the functional integration over the connection, viewed as an independent variable, is Gaussian and gives a functional integral with the action purely in terms of the tetrad type variables. In the discrete framework, such an integration over the connection gives a certain phase of the result and a certain module of the result.

For the module, it is appropriate to use an expansion over the discrete analogs of the Arnowitt-Deser-Misner \cite{ADM,ADM1} lapse-shift functions $(N, N^i)$, for a nontrivial module appears already in the zero order of this expansion. In the continuum GR, $(N, N^i)$ are the tetrad 4-vectors with the temporal world vector index (covariant). They enter the tetrad-connection (Cartan-Weyl) GR action linearly, through the timelike (with spatio-temporal indices) bi-vectors. The discrete $(N, N^i)$ are the 4-vectors of certain ("timelike") edges, and their scales enter the tetrad-connection Regge action still linearly, through the tensors of certain ("timelike") triangles. Each term of the expansion over $(N, N^i)$ (as for zero $(N, N^i)$) factorizes over "spacelike" triangles (analogs of the continuum bi-vectors with spatio-spatial world covariant indices) and is calculable. The resulting module or measure in terms of the purely tetrad type variables has a pronounced maximum at the values of the areas of the spacelike triangles defined by the Planck scale and some parameter ($\eta$) characterizing a certain freedom in constructing the functional measure. We can say that their areas or the lengths of the spacelike edges are loosely fixed dynamically, while the timelike edge vectors $(N, N^i)$ can be manually set (like fixing the gauge in the continuum case) as defining the mutual position of any two neighboring simplicial spatial 3D leaves of a similar structure constituting the simplicial space-time.

For the phase, it is appropriate to use the stationary phase expansion, for a nonzero phase appears already in the zero order - it is exactly the Regge action $S ( \rl )$ in terms of the edge lengths $\rl$ since classically excluding the connection from the tetrad-connection Regge action gives just $S ( \rl )$.

The above said can be resumed by the following symbol for the functional integral in terms of the purely length type variables obtained by the functional integration over the connection type variables,
\begin{equation}\label{FexpiS(l)}                                           
\int \exp [i S ( \rl ) ] F ( \rl ) D \rl ,
\end{equation}

\noindent $D \rl = \prod_k \d l_k$ is the collective Lebesgue measure, $\rl = (l_1, \dots, l_n )$ is the set of the edge lengths. The Regge action $S ( \rl )$ appears here as the leading term in the stationary phase expansion for the phase, and in $F ( \rl )$ we take the leading term in the expansion over the discrete $(N, N^i)$ for the module. $F ( \rl )$ turns out to have a maximum at the areas of the spacelike triangles being $a^2 / 2$, where the length scale
\begin{equation}\label{a=sqrt}                                              
a = \sqrt{ 32 G ( \eta - 5) / 3 } .
\end{equation}

\noindent Here $\eta$ is a parameter of the theory defining the 4-simplex volume degree factors $V^\eta$ in the functional measure analogous to the metric determinant factors $(-g)^{\eta / 2}$ in the continuum measure. (The continuum measure is determined up to such a factor, which leads, eg, to the measures of Misner \cite{Mis} or DeWitt \cite{DeW}.) This loosely fixes the spacelike edge lengths at the scale $a$. Both these expansions, for the phase and for the module, are consistent and allow to limit ourselves by their leading terms for a sufficiently large $a$ and thus $\eta$.

To formulate this in a more quantitative manner, we implicitly pass from $\rl$ to some new collective variable $u = (u_1, \dots, u_n )$ such that $F( \rl ) D \rl = D u$ is the Lebesgue measure. Besides, we can choose some stationary collective point $\rl_0 = (l_{01}, \dots, l_{0n} )$, that is, ensuring the validity of the equations of motion (Regge equations) in it,
\begin{equation}\label{dS/dl=0}                                             
\frac{\partial S(l_0)}{\partial l} = 0.
\end{equation}

\noindent Then the Taylor expansion of $S(\rl)$ around $\rl_0=\rl(u_0)$, $u_0 = (u_{01}, \dots, u_{0n} )$, over $\Delta u = u - u_0$ begins with a second-order term,
\begin{equation}\label{Sdudu}                                               
S (\rl ) = \frac{1}{2} \sum_{j, k, l, m} \frac{\partial^2 S (\rl_0 )}{\partial l_j \partial l_l} \frac{\partial l_j (u_0 )}{\partial u_k} \frac{\partial l_l (u_0 )}{\partial u_m} \Delta u_k \Delta  u_m + \dots .
\end{equation}

\noindent The equations of motion (\ref{dS/dl=0}), that is, the requirement of an extremum of the zero order term $S(\rl_0)$, are insufficient for fixing all the variables $l_k$. The latter, in particular, their scale, are loosely fixed by the requirement of an extremum (maximum) of the contribution of the second order term, that is, of the minimum of the determinant of the bilinear form over $\Delta u$ in (\ref{Sdudu}) or the maximum of
\begin{equation}\label{def-l0}                                              
F (\rl_0 )^2 \det \left \| \frac{\partial^2 S (\rl_0 )}{\partial l_i \partial l_k} \right \|^{-1}.
\end{equation}

\noindent The matrix $\partial^2 S (\rl_0 ) / \partial l_i \partial l_k$ has zero order in the scale of edge lengths. Geometrically, the edge length scale can not change rapidly from simplex to simplex, and the matrix $\partial^2 S (\rl_0 ) / \partial l_i \partial l_k$ is just close to a diagonal one (only those $l_i$ and $l_k$ "interact" in the Regge action $S$ which refer to the same 4-simplex). Therefore, it is expected that the inclusion of the determinant of this matrix in (\ref{def-l0}) will not lead to an essential change in the extreme point $\rl_0$ of (\ref{def-l0}) compared to the maximum of only $F (\rl_0 )$. This also means some sufficient uniformity of the elementary length scale.

In principle, an elementary length scale could be defined, eg, by finding length vacuum expectations. In the above definition of the length scale $\rl_0$, the perturbative interaction with gravitons acts as a probe, which seems quite natural.

In the paper \cite{our2}, we considered calculating the Regge action in terms of the simplicial metric resembling calculating the Hilbert-Einstein action through intermediate finding the Christoffel symbols. Discrete Christoffel symbols or affine connection matrices were used there, and an exact expression of the Regge action in terms of the piecewise constant metric was considered for the general simplicial structure. A particular periodic simplicial structure with the 4-cube cell divided by the diagonals into 24 4-simplices was considered. For this structure, we found that the Regge action in the considered form, arranged in a series over metric variations between the {\it 4-simplices}, is in the leading order a finite-difference form of the Hilbert-Einstein action in terms of the metric variations between the {\it 4-cubes},
\begin{eqnarray}\label{DM+MM}                                               
\hspace{-5mm}
\sum_{\rm 4-cubes} \K^{\lambda \mu}_{~~~ \lambda \mu} \sqrt{g} , \mbox{ where } \K^\lambda_{~ \, \mu \nu \rho} = \Delta_\nu M^\lambda_{\rho \mu} - \Delta_\rho M^\lambda_{\nu \mu} + M^\lambda_{\nu \sigma} M^\sigma_{\rho \mu} - M^\lambda_{\rho \sigma} M^\sigma_{\nu \mu} , \nonumber \\
M^\lambda_{\mu \nu} = \frac{1}{2} g^{\lambda \rho} (\Delta_\nu g_{\mu \rho} + \Delta_\mu g_{\rho \nu} - \Delta_\rho g_{\mu \nu}), ~~~ \Delta_\lambda = 1 - \bT_\lambda .
\end{eqnarray}

\noindent $\K^\lambda_{~ \, \mu \nu \rho}$ is the finite-difference form of the Riemannian tensor $R^\lambda_{~ \, \mu \nu \rho}$, $T_\lambda$ (respectively, Hermitian conjugate $\bT_\lambda$) is the shift operator along the edge $\lambda$ or the coordinate $x^\lambda$ in the forward (respectively, backward) direction. Here, the coordinate steps are equal to 1 and can be understandably generalized to values other than 1.

Thereby, the analysis of the Regge skeleton equations is reduced to the analysis of the finite-difference Einstein equations (with the possibility of a regular consideration of the further corrections).

The use of a periodic simplicial structure respects the above mentioned uniformity of the elementary length scale and makes the most appropriate that the metric ansatz, substituted into the action (\ref{DM+MM}), be formulated in Cartesian type coordinates without requiring a priori spherical symmetry. It turns out that such an ansatz should cover the case of the Painlev\'{e}-Gullstrand metric in such coordinates, and it is convenient to take the 3+1 ADM form of the metric for it. Before the substitution, we should convert the ansatz into a form in which $N, N^k$, as mentioned, are constant parameters.

In Section \ref{general}, the metric ansatz is transformed to that analogous to the Lemaitre metric (as a simplest example with fixed $N = 1, N^k = 0$), and also the region of sufficiency of the leading order over metric variations between the 4-simplices is considered from different viewpoints. In Section \ref{leading}, the leading order over metric variations is considered. In this order, the finite-difference form of the GR action on the metric ansatz transformed to the Lemaitre type coordinates coincides with this form on the original ansatz in the Painlev\'{e}-Gullstrand type coordinates due to invariance. The Einstein equations (in the ADM formalism), the discrete form of them and of the Painlev\'{e}-Gullstrand solution that follows, the metric, discrete Riemann tensor and Kretschmann scalar at the center are considered.

\section{General form of the metric}\label{general}

Now we consider a black hole type solution of the equations of motion (\ref{dS/dl=0}). In accordance with the above mentioned mechanism for fixing the edge lengths, the spacelike edge lengths are loosely fixed dynamically while the discrete lapse-shift vectors are given as parameters. A particular case is (the discrete analog of) the synchronous frame $N = 1$, $N^i = 0$. For the Schwarzschild black hole, this means using (a discrete analog of) the Lemaitre type metric \cite{Lemaitre}, which we write in the form using a radial type coordinate $r_1$ such that at $r_g = 0$ it be the standard $r$ \cite{Stan},
\begin{eqnarray}\label{Lem}                                                 
& & \d s^2 = - \d \tau^2 + \frac{r_1}{r(r_1, \tau )} \d r^2_1 + r^2 (r_1 , \tau ) \d \Omega^2 \nonumber \\ & & = - \d \tau^2 + ( \d r(r_1, \tau ) |_{\tau = const} )^2 + r^2 (r_1 , \tau ) \d \Omega^2 , ~~~ r^{3/2} = r^{3/2}_1 - \frac{3}{2} \sqrt{r_g} \tau .
\end{eqnarray}

\noindent We see that the 3D sections $\tau = const$ possess the flat metric. We are interested in a finite-difference analogue of the equations which lead to (\ref{Lem}). For a periodic simplicial structure and a certain length scale such an analog can be written in Cartesian type coordinates,
\begin{eqnarray}                                                            
& & \by = r_1 \bn , ~~~ \bn^2 \equiv \sum_k n^k n^k = 1 , ~~~ \d \Omega^2 = \d \bn^2 , \nonumber \\ & & \d s^2 = - \d \tau^2 + \frac{r^2}{r^2_1} \d \by^2 + \left ( \frac{r_1}{r} - \frac{r^2}{r^2_1} \right ) \frac{(\by \d \by)^2}{r^2_1} .
\end{eqnarray}

Taking a step back in obtaining (\ref{Lem}), we return from $r_1$ to the original $r$ and get the Painlev\'{e}-Gullstrand metric \cite{Painleve,Gullstrand},
\begin{equation}\label{Pan}                                                 
\d s^2 = - \d \tau^2 + \left ( \d r + \sqrt{\frac{r_g}{r}} \d \tau \right )^2 + r^2 \d \Omega^2 .
\end{equation}

\noindent The 3D sections $\tau = const$ possess the flat metric here too.

According to the above said, we would like to write down the Einstein equations in a finite-difference form and re-solve the problem with this input. Any Regge manifold does not possess the spherical symmetry. The latter is restored when averaging over possible simplicial structures. As for a given structure, the criterium for the solution might be that at large distances, where the metric variations between the 4-simplices are small and the Einstein equations in a finite-difference form are close to their continuum form, the solution of interest should be close to (\ref{Lem}).

\begin{figure}[h]
\unitlength 1pt
\begin{picture}(120,152)(-140,-62)
\put(-43,-43){\line(-1,-1){12}}
\put(-43,-43){\line(1,0){160}}
\put(-43,-43){\line(0,1){130}}
\put(-43,-43){\line(1,1){130}}
\put(110,-56){$r_1^{3/2}$}
\put(65,-56){$(4a)^{3/2}$}
\put(23,-56){$(3a)^{3/2}$}
\put(-13,-56){$(2a)^{3/2}$}
\put(-34,-56){$a^{3/2}$}
\put(-45,-56){$0$}
\put(-18,6){$r=0$}
\put(-23,20){$r=a$}
\put(-25,19){\line(1,0){24}}
\put(-1,19){\vector(2,-1){23}}
\put(-19,29){$r=2a$}
\put(-21,28){\line(1,0){29}}
\put(8,28){\vector(2,-1){42}}
\put(-75,75){$\frac{3}{2} \tau \sqrt{r_g}$}
\put(-28,-43){\line(0,1){15}}
\put(-28,-43){\line(2,1){27}}
\put(-28,-28){\line(2,1){27}}
\put(-1,-43){\line(0,1){42}}
\put(-1,-43){\line(5,2){36}}
\put(-1,-28){\line(3,1){36}}
\put(-1,-14.5){\line(3,1){36}}
\put(-1,-1){\line(2,1){36}}
\put(35,-43){\line(0,1){78}}
\put(35,-43){\line(3,1){42}}
\put(35,-29){\line(3,1){42}}
\put(35,-14.5){\line(3,1){42}}
\put(35,-1){\line(5,2){42}}
\put(35,17){\line(5,2){42}}
\put(35,35){\line(3,1){42}}
\put(22,35){$A_n$}
\put(25,49){$A_{n+1}$}
\put(77,-43){\line(0,1){120}}
\put(77,-44){\line(3,1){40}}
\put(77,-28){\line(4,1){40}}
\put(77,-14.5){\line(4,1){40}}
\put(77,-1){\line(3,1){40}}
\put(77,17){\line(3,1){40}}
\put(77,35){\line(4,1){40}}
\put(77,49){\line(4,1){40}}
\put(77,63){\line(4,1){40}}
\put(77,77){\line(3,1){30}}
\put(-28,-28){\line(1,0){145}}
\put(-1,-1){\line(1,0){118}}
\put(-14.5,-14.5){\line(1,0){131.5}}
\put(35,35){\line(1,0){82}}
\put(17,17){\line(1,0){100}}
\put(77,77){\line(1,0){40}}
\put(63,63){\line(1,0){54}}
\put(49,49){\line(1,0){68}}
\put(49,49){\line(2,1){28}}
\put(-33,-48){\line(1,1){15}}
\put(-12,-27){\line(1,1){10}}
\put(4,-11){\line(1,1){10}}
\put(20,5){\line(1,1){10}}
\put(36,21){\line(1,1){10}}
\put(52,37){\line(1,1){10}}
\put(68,53){\line(1,1){10}}
\put(84,69){\line(1,1){10}}
\put(-6,-48){\line(1,1){15}}
\put(15,-27){\line(1,1){10}}
\put(31,-11){\line(1,1){10}}
\put(47,5){\line(1,1){10}}
\put(63,21){\line(1,1){10}}
\put(79,37){\line(1,1){10}}
\put(95,53){\line(1,1){10}}
\put(111,69){\line(1,1){10}}
\put(30,-48){\line(1,1){15}}
\put(51,-27){\line(1,1){10}}
\put(67,-11){\line(1,1){10}}
\put(83,5){\line(1,1){10}}
\put(99,21){\line(1,1){10}}
\put(115,37){\line(1,1){10}}
\put(127,42){$r=3a$}
\put(72,-48){\line(1,1){15}}
\put(93,-27){\line(1,1){10}}
\put(109,-11){\line(1,1){15}}
\put(127,1){$r=4a$}
\end{picture}
\caption{A triangulation in the neighborhood of $r = 0$ in the Lemaitre type coordinates. $A_n A_{n+1}$ is an edge at $r=0$. Because of the events of ending worldlines at $r = 0$, intervals $\Delta \tau$ between neighboring 3D sections $\tau = const$ look non-uniform, but another variant is possible in which $\Delta \tau$ are constant, but the (hyper)cubic cells are distorted in the vicinity of $r = 0$.}\label{triangulation}
\end{figure}
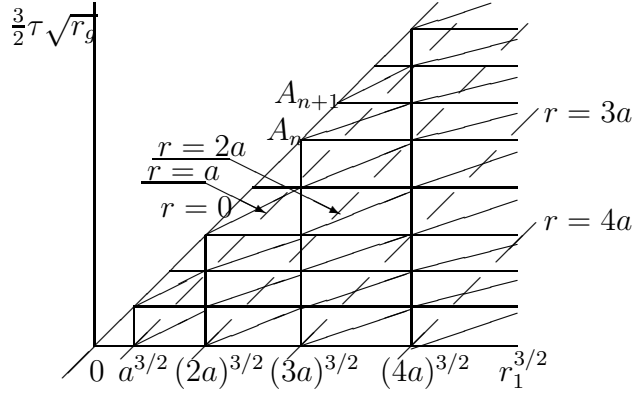

The criterium for retaining only the leading order over metric variations between the 4-simplices in the above mentioned expansion over such variations is smallness of these variations. When we go to a smaller distances, using the leading order over metric variations between the 4-simplices may become insufficient. Using the suggested required matching between the finite-difference and continuum solutions, we can trace when the metric variations between the 4-simplices can not be small. To this end, we can consider a simplicial structure with 4-cube cells, whose spacelike bases are in the (flat) 3D sections $\tau = const$, timelike edges are geodesic lines $r_1 = const$, orthogonal to these sections, and their timelike length is $\Delta \tau$ (the difference between the neighboring sections $\tau = const$), Fig.~\ref{triangulation}.

If a certain line $r_1 = const$ corresponds to $r > 0$ at $\tau = \tau_0$, then it reaches the singularity $r = 0$ at $\tau = \tau_0 + T$. This event means a defect violating the regular coverage of the space-time by the considered orthogonal lattice. The condition that the number of such defects (=1) be negligible compared to the number of the regular vertices along this geodesic ($T / \Delta \tau$) is $T \gg \Delta \tau$. Fixing $\Delta \tau$ similarly to fixing the lapse function $N$ contributes to choosing a certain measurement procedure. Having the dynamically fixed elementary length scale $a$, the choice $\Delta \tau \ll a$ means an excessive accuracy, $\Delta \tau \gg a$ lack of accuracy, and the essential deviation of $\Delta \tau / a$ from unity in both these cases makes the description of the system somewhat singular. Like averaging over possible simplicial structures, averaging over different timelike lengths seems to be appropriate, but now we can take for estimate $\Delta \tau \simeq a$, which also reflects a symmetry between space and time. Thus, at
\begin{equation}\label{T>>Dtau}                                            
T = \frac{2r^{3 / 2}}{3 \sqrt{r_g}} \gg \Delta \tau \simeq a \mbox{ or } r_g \ll \frac{r^3 }{a^2 } ,
\end{equation}

\noindent metric variations between the 4-simplices are small. This is a kind of a nonlocal consideration. In a more local approach, since we are interested in a finite-difference form of differential equations, it is sufficient to take $\tau = 0$ and a few first 3D sections after $\tau = 0$ (enough to form a 4-geometry), that is, $\tau \simeq a$. The metric from $\tau = 0$ to $\tau \simeq a$ is distorted from being flat, and the measure of this distortion is closeness of $r / r_1$ to unity,
\begin{equation}\label{Dr<<r1}                                             
\left | \frac{r_1 (r, \tau ) - r}{r_1 (r, \tau )} \right | \ll 1 \mbox{ at } r_g \ll \frac{r^3 }{a^2 } .
\end{equation}

\noindent The same follows from the typical curvature in the continuum GR $R \sim r_g / r^3$ (from the curvature invariants). For the elementary area scale $a^2$, this gives a typical value of the angle defect $\alpha \sim R a^2$ and the condition for its smallness,
\begin{equation}                                                           
\alpha \sim a^2 R \sim a^2 \frac{r_g}{r^3} \ll 1 \mbox{ at } r_g \ll \frac{r^3 }{a^2 } .
\end{equation}

In (\ref{DM+MM}), the metric in the 4-simplices/cubes enters, and these values can be viewed as particular values of some smooth (interpolating) field $g_{\lambda \mu}$. In the zeroth order over metric variations between the 4-simplices, the Lemaitre metric (\ref{Lem}) could be taken as $g_{\lambda \mu}$. Having in view the minimal nonzero $r = a$ of a vertex, we get from the above estimates that the Lemaitre $g_{\lambda \mu}$ can be prolonged to $r \gsimeqscr a$ at
\begin{equation}\label{r_g<<a}                                             
r_g \ll a ,
\end{equation}

\noindent though this is not physically quite an interesting case (there is no horizon as such). Note that $a \gg 1$ at $\eta \gg 1$ (see (\ref{a=sqrt})), and $r_g \gg 1$ admits (\ref{r_g<<a}).

In non-leading orders, generally speaking, it should be a common synchronous metric. To bring it into the form most adapted to the task at hand, it is convenient to issue from the Painlev\'{e}-Gullstrand metric (\ref{Pan}), rewritten in Cartesian type coordinates,
\begin{equation}\label{PanCartesian}                                       
\d s^2 = - \d \tau^2 + \sum^3_{k = 1} (\d x^k + f^k \d \tau)^2 , ~~~ f^k = \sqrt{\frac{r_g }{r}} \frac{x^k}{r} .
\end{equation}

\noindent This can be naturally generalized to the 3+1 ADM form of metric \cite{ADM1} ,
\begin{equation}\label{3+1}                                                
\d s^2 = - ( N \d \tau )^2 + g_{k l} (\d x^k + f^k \d \tau ) (\d x^l + f^l \d \tau ) .
\end{equation}

\noindent Here $k, l, \dots = 1, 2, 3$; in the leading order over metric variations or in the continuum limit $f^k$ are given by (\ref{PanCartesian}), $N = 1$, $g_{kl} = \delta_{kl}$.

We can perform the change of variables in (\ref{3+1}) which generalizes the transition from the Painlev\'{e}-Gullstrand (\ref{Pan}) to the Lemaitre (\ref{Lem}) metric. Finding this change of variables $( \bx , \tau ) \to ( \by , \tau )$, $x^k = x^k ( \by , \tau )$ amounts to solving the differential equations
\begin{eqnarray}\label{dx/dt+f=0}                                       
& & \frac{\partial x^k ( \by , \tau )}{\partial \tau } + f^k ( \bx (\by , \tau ) , \tau ) = 0 , ~~~ x^k (\by , 0 ) = y^k \\ & & \mbox{ or } ~~~ x^k ( \by , \tau ) = y^k - \int^\tau_0 f^k ( \bx (\by , \tau ) , \tau ) \d \tau
\end{eqnarray}

\noindent in the integral form. Then the metric reads
\begin{equation}\label{likeLemaitre}                                       
\d s^2 = - ( N \d \tau )^2 + g_{k l} \frac{\partial x^k }{\partial y^m } \frac{\partial x^l }{\partial y^n } \d y^m \d y^n .
\end{equation}

\noindent In principle, the task of replacing variables could be complicated by the requirement of reducing $N$ to unity, that is, obtaining a completely synchronous reference system as a result. But since $N = 1$ in the leading order over metric variations, we expect it is bounded on the whole space-time and its possible dependence on the coordinates  will not affect convergence properties of the expansion over lapse-shift functions $(N, N^i)$ (here $(N, 0)$) for the above functional integral measure $F ( \rl )$ in (\ref{FexpiS(l)}). In the specific calculations of the present paper, in the leading order over metric variations, a Lemaitre type metric is used only intermediately (only the fact of its existence is important), and the specific value of $N$ does not manifest itself.

We can represent the solution of equations (\ref{dx/dt+f=0}) as a combination of a certain (expectedly large) part of the transformation $( \bx , \tau ) \to ( \by , \tau )$, expressed in quadratures, and other less significant transformations.

For that, we separate the radial and angle parts of $\bx$: $\bx = r \bn$ (with respect to the 3-metric $g_{k l} = \delta_{k l}$: $\sum_k n^k n^k = 1$). First, we take the longitudinal part of equations (\ref{dx/dt+f=0}),
\begin{equation}                                                           
\frac{\partial r}{\partial \tau} + \vf \cdot \bn = 0 .
\end{equation}

\noindent In the case of the Painlev\'{e}-Gullstrand metric, solving this via
\begin{equation}\label{int-r-r1-dr/fn+tau}                                 
\int^r_{r_1} \frac{\d r}{\vf \cdot \bn} + \tau = 0
\end{equation}

\noindent introduces the Lemaitre variable $r_1$ as a function of $r$ and $\tau$. Now consider the general case when $\vf \cdot \bn$ is a function of $r, \bn , \tau$. Then (\ref{int-r-r1-dr/fn+tau}) defines a function $r = r(r_1 , \bn , \tau )$, which, as a rule, does not vanish the spatio-temporal components of the metric, but can serve as a substitution, after which it remains to go from $r_1 , \bn$ to some $\tlr_1 , \tbn$, close to $r_1 , \bn$, if the metric is close to the Painlev\'{e}-Gullstrand one. We write out the differential 1-form included in the metric 2-form (\ref{3+1}),
\begin{eqnarray}\label{dx+fdt}                                             
& & \d \bx + \vf \d \tau = \bn \d r + r \d \bn + \vf \d \tau = \bn \left ( \frac{\partial r}{\partial r_1} \right )_{\bn, \ttau} \d r_1 + r \d \bn \nonumber \\ & & + \bn \left ( \left ( \frac{\partial r}{\partial \bn} \right )_{\trone , \ttau } \! \! \! \cdot \d \bn \right ) + \left ( \bn \left ( \frac{\partial r}{\partial \tau} \right )_{\trone , \bn} \! \! \! + \vf \right ) \d \tau .
\end{eqnarray}

\noindent The derivatives of $r$ over $r_1 , \bn , \tau$ are found by differentiating (\ref{int-r-r1-dr/fn+tau}),
\begin{eqnarray}                                                           
& & \left ( \frac{\partial r}{\partial r_1} \right )_{\bn, \ttau} = \frac{(\vf \cdot \bn)(r, \bn , \tau)}{(\vf \cdot \bn)(r_1, \bn , \tau)} , \left ( \frac{\partial r}{\partial \bn} \right )_{\trone , \ttau } = (\vf \cdot \bn)(r, \bn , \tau) \frac{\partial }{\partial \bn} \int^{r_1}_r \frac{\d r}{\vf \cdot \bn} , \nonumber \\ & & \left ( \frac{\partial r}{\partial \tau} \right )_{\trone , \bn} = (\vf \cdot \bn)(r, \bn , \tau) \left ( \frac{\partial}{\partial \tau} \int^{r_1}_r \frac{\d r}{\vf \cdot \bn} - 1 \right ) .
\end{eqnarray}

\noindent Now we pass from $r_1 $, $\bn$ to some $\tlr_1 $, $\tbn$ and admit that this transformation $r_1 = r_1 (\tlr_1 , \tbn , \tau ) , \bn = \bn (\tlr_1 , \tbn ,$ $\! \! \! $ $\tau )$ (weakly) depends on $\tau$ to remove the residual $\d \tau$ term in (\ref{dx+fdt}). Requiring vanishing this term, we get the equations for this transformation,
\begin{eqnarray}\label{dr1/dt,dn/dt}                                       
& & \frac{\partial r_1}{\partial \tau} = (\vf \cdot \bn)(r_1, \bn , \tau) \left ( \frac{1}{r} \vf_\perp \cdot \frac{\partial }{\partial \bn} - \frac{\partial}{\partial \tau} \right ) \int^{r_1}_r \frac{\d r}{\vf \cdot \bn} , ~~~ \frac{\partial \bn}{\partial \tau} = - \frac{\vf_\perp}{r} , \nonumber \\ & & \vf_\perp \equiv \vf - \bn (\vf \cdot \bn), ~~~ r_1 |_{\textstyle \tau = 0} = \tlr_1 , ~~~ \bn |_{\textstyle \tau = 0} = \tbn .
\end{eqnarray}

\noindent Then the 1-form (\ref{dx+fdt}) reads
\begin{equation}                                                           
\bn \frac{(\vf \cdot \bn)(r, \bn , \tau)}{(\vf \cdot \bn)(r_1, \bn , \tau)} \d r_1 |_\tau + r \d \bn |_\tau + \bn (\vf \cdot \bn)(r, \bn , \tau) \left ( \d \bn |_\tau \cdot \frac{\partial }{\partial \bn} \int^{r_1}_r \frac{\d r}{\vf \cdot \bn} \right ),
\end{equation}

\noindent where $|_\tau$ means $\tau = const$,
\begin{equation}\label{|_tau}                                              
\d r_1 |_\tau = \frac{\partial r_1}{\partial \tlr_1} \d \tlr_1 + \frac{\partial r_1}{\partial \tbn} \d \tbn , ~~~ \d n^k |_\tau = \frac{\partial n^k}{\partial \tlr_1} \d \tlr_1 + \frac{\partial n^k}{\partial \tbn} \d \tbn .
\end{equation}

\noindent Finally, we pass to Cartesian type coordinates
\begin{equation}\label{y=rn}                                               
\by = \tlr_1 \tbn , ~~~ \tlr_1 = \sqrt{\by^2} , ~~~ \d \tlr_1 = \tlr_1^{-1} \by \cdot \d \by , ~~~ \d \tbn = \tlr_1^{-1} \d \by - \tlr_1^{-3} \by (\by \cdot \d \by) .
\end{equation}

\noindent The metric is
\begin{eqnarray}\label{genLem}                                             
& & \d s^2 = - ( N \d \tau )^2 + \left ( g_{k l} n^k n^l \right ) \left [ \frac{(\vf \cdot \bn)(r, \bn , \tau)}{(\vf \cdot \bn)(r_1, \bn , \tau)} \right ]^2 \left [ \d r_1 |_\tau \phantom{\int^1_0} \right. \nonumber \\ & & \left. + (\vf \cdot \bn)(r_1, \bn , \tau) \left ( \d \bn |_\tau \cdot \frac{\partial }{\partial \bn} \int^{r_1}_r \frac{\d r}{\vf \cdot \bn} \right ) \right ]^2 \nonumber \\ & & + 2 r \frac{(\vf \cdot \bn)(r, \bn , \tau)}{(\vf \cdot \bn)(r_1, \bn , \tau)} \left ( g_{k l} n^k \d n^l \right ) \left [ \d r_1 |_\tau \phantom{\int^1_0} \right. \nonumber \\ & & \left. + (\vf \cdot \bn)(r_1, \bn , \tau) \left ( \d \bn |_\tau \cdot \frac{\partial }{\partial \bn} \int^{r_1}_r \frac{\d r}{\vf \cdot \bn} \right ) \right ] + r^2 g_{k l} \d n^k \d n^l .
\end{eqnarray}

\noindent Together with (\ref{int-r-r1-dr/fn+tau}), (\ref{dr1/dt,dn/dt}), (\ref{|_tau}), (\ref{y=rn}), this is an exact expression generalizing the Le\-mai\-tre form for an arbitrary metric (\ref{3+1}), a more detailed formula (\ref{likeLemaitre}) where the leading and non-leading contributions are singled out if the original metric is close to the Painlev\'{e}-Gullstrand one. In particular, $\vf_\perp$ and $\partial (\vf \cdot \bn) / \partial \bn $ are non-leading in this case, therefore, at $\partial \vf / \partial \tau = 0$, the equations for the transformation (\ref{dr1/dt,dn/dt}) give $\partial r_1 / \partial \tau$ being of the second order over deviation from the Painlev\'{e}-Gullstrand metric. Also $g_{k l} n^k \d n^l = 0$ in $\d s^2$ (\ref{genLem}) in the leading order and so on. In the discrete framework, this deviation will be characterized by the metric variations from simplex to simplex or finite differences $\Delta$ .

\section{Field equations in the leading order over metric variations}\label{leading}

We can substitute the metric ansatz (\ref{likeLemaitre}) or, in more detail, (\ref{genLem}), into the finite-difference form of the GR action (\ref{DM+MM}). The finite differences approximately obey the rules for the derivatives, in particular, the chain rule for computing the derivative of the composition of two functions, for example,
\begin{equation}\label{chain}                                              
\frac{\Delta g_{kl} (\bx (\by , \tau ) , \tau )}{\Delta y^m} = \frac{\Delta x^n ( \by , \tau )}{\Delta y^m} \frac{\Delta g_{kl} (\bx (\by , \tau ) , \tau )}{\Delta x^n (\by , \tau )} ,
\end{equation}

\noindent and the product rule. The condition of validity of these properties of the finite differences of the coordinate change expressions is the same as the condition of smallness of the metric variations, if we apply it to that point involved in any considered finite difference which has a smaller $r$; for example, taking the radial coordinates $r$ and $r_1$ (which are certain functions of the Cartesian type coordinates $x^k$, $y^k$), we have for the accuracy of reproducing the typical coordinate derivative by the corresponding finite difference,
\begin{equation}\label{Dr1-dr1<<1}                                         
\left | \frac{a}{r_1 (r + a, \tau ) - r_1(r, \tau )} - \frac{\partial r}{\partial r_1(r + a, \tau )} \right | \left | \frac{\partial r}{\partial r_1(r + a, \tau )} \right |^{-1} \ll 1 \mbox{ at } r_g \ll \frac{r^3 }{a^2 } ,
\end{equation}

\noindent like (\ref{Dr<<r1}). It is also taken into account that the minimal nonzero $r=a$ of a vertex in a 3D section. Here we take into account that it is $r$ that measures the radial lengths in the sections $\tau = const$, and not $r_1$ (since $r^{-1} r_1 \d r^2_1 = \d r^2 |_{\tau = const}$ in $\d s^2$ (\ref{Lem})), although if (\ref{Dr1-dr1<<1}) holds, there is no difference, $\Delta r$ or $\Delta r_1$ has the scale $a$.

If we consider the leading order over metric variations when operating with the finite differences in the form for the action, we can handle these differences as the derivatives and, in particular, use the general covariance, reducing this form from the coordinates $y^k$ back to $x^k$ when substituting the metric ansatz (\ref{likeLemaitre}) into the action. Thereby we get a finite-difference form of the action with the 3+1 ADM form of metric (\ref{3+1}). Note that such a return to the (finite-difference) action on the metric (\ref{3+1}) takes place only in the leading order over metric variations. In non-leading orders, in the equations like (\ref{chain}), the corrections of the higher order over $\Delta$ should be taken into account in addition to the main terms in which the differences are replaced by the corresponding derivatives. (Remind that the procedure of loose fixing the elementary edge lengths implies the constant discrete lapse-shift functions and thus a construction of the 4D space-time of the 3D leaves analogous to that at $(N, N^k) = const$ in the continuum case or, in particular, in a synchronous frame; that is why do we need to operate beginning from the Lemaitre like frame.)

Thus, we are interested in (the finite-difference form of) the action, a bulk expression of which (up to surface terms) takes the form \cite{ADM1,MisTorWhe} (hereinafter, the coefficient $(16 \pi G )^{-1}$ is suppressed),
\begin{eqnarray}                                                        
& & \hspace{-9mm} S = \int \left \{ - g_{k l} \frac{\partial \pi^{k l}}{\partial \tau} + N \sqrt{g} \left [ ^3 \! R + g^{-1} \left ( \frac{1}{2} \pi^k{}_k \pi^l{}_l - \pi^k{}_l \pi^l{}_k \right ) \right ] + 2 f_k \pi^{k l}{}_{| l} \right \} \d^3 x \d \tau , \\ & & \label{pi=dg/dt} \hspace{-5mm} \pi_{k l} = \sqrt{g} (g_{k l} K^m{}_m - K_{k l}) , ~~~ K_{k l} = \frac{1}{2N} \left ( f_{k | l} + f_{l | k} - \frac{\partial g_{k l}}{\partial \tau} \right ) .
\end{eqnarray}

\noindent Raising and lowering Latin indices is done using the 3D metric $g_{k l}$, the symbol $|$ as an index means the covariant derivative, and $^3 \! R$ is the curvature scalar for this metric. Varying $S$ gives
\begin{eqnarray}                                                     
& & \frac{\delta S}{\delta N} = \sqrt{g} \left [ ^3 \! R + g^{-1} \left ( \frac{1}{2} \pi^k{}_k \pi^l{}_l - \pi^k{}_l \pi^l{}_k \right ) \right ] , \\ & &
\frac{\delta S}{\delta f_k} = 2 \pi^{k l}{}_{| l} , \\ & &
\frac{\delta S}{\delta g_{k l}} = -N \sqrt{g} \left ( ^3 \! R^{k l} - \frac{1}{2} g^{k l} \, {}^3 \! R \right ) - \frac{1}{2} \frac{N}{\sqrt{g}} g^{k l} \left ( \frac{1}{2} \pi^m{}_m \pi^n{}_n - \pi^m{}_n \pi^n{}_m \right ) \nonumber \\ & & + 2 \frac{N}{\sqrt{g}} \left ( \frac{1}{2} \pi^m{}_m \pi^{k l} - \pi^k{}_m \pi^{m l} \right ) - \frac{\partial \pi^{k l}}{\partial \tau} + \sqrt{g} (N^{| k l} - g^{k l} N^{| m}{}_{| m}) \nonumber \\ & & + (\pi^{k l} f^m)_{| m} - f^k{}_{| m} \pi^{m l} - f^l{}_{| m} \pi^{m k} .
\end{eqnarray}

\noindent Equating these expressions to zero (in empty spacetime) gives combinations of the Einstein equations, and $\delta S / \delta N = 0$ and $\delta S / \delta f_k = 0$ are the equations for initial conditions in the Hamiltonian formalism, and $\delta S / \delta g_{k l} = 0$ (and analogous one defining $\pi_{k l}$ in terms of $\partial g_{k l} / \partial \tau$ (\ref{pi=dg/dt})) are the dynamical equations. In the Lagrangian formalism in our case and especially in the stationary problem ($\partial \pi_{k l} / \partial \tau = 0$, $\partial g_{k l} / \partial \tau = 0$), we can look at these equations from a slightly different angle. Namely, $\delta S / \delta g_{k l} = 0$ can be considered as those defining 3D geometry ($^3 \! R_{k l}$ and, therefore, $^3 \! R_{klmn}$) from knowing $N$, $f_k$. $\delta S / \delta f_k = 0$ can be viewed as specifying $f_k$. $\delta S / \delta N = 0$ looks as a condition on $^3 \! R_{k l}$ or $^3 \! R_{klmn}$. Since, however, the 3D curvature is already defined by $\delta S / \delta g_{k l} = 0$, the term $^3 \! R$ can be excluded from $\delta S / \delta N = 0$. This is achieved by forming a combination of $\delta S / \delta N = 0$ and the trace of $\delta S / \delta g_{k l} = 0$. The resulting equation and $\delta S / \delta f_k = 0$ form a system for $N$, $f_k$. To get a more compact dependence on $f_k$, it turns out to be appropriate to take a combination of this equation and $\delta S / \delta f^k = 0$,
\begin{equation}\label{dS/dg-dS/df-dS/dN}                                  
0 = \! \frac{N}{\sqrt{g}} \left [ \frac{ g_{k l}}{2} \frac{\delta S}{\delta g_{k l}} - \! f^k \frac{\delta S}{\delta f^k} - \! \frac{ N}{4} \frac{\delta S}{\delta N} \right ] \! = \! \frac{1}{2} (f^k f_k)^{| l}_{| l} - \! \frac{1}{4} (f^{k | l} \! - \! f^{l | k} )(f_{k | l} \! - \! f_{l | k} ) + \dots ,
\end{equation}

\noindent so that
\begin{eqnarray}\label{dd(ff)=}                                            
& & \hspace{-5mm} \frac{1}{2} (f^k f_k)^{| l}_{| l} - \frac{1}{4} (f^{k | l} - f^{l | k} )(f_{k | l} - f_{l | k} ) = \nonumber \\ & & \hspace{-5mm} - ^3 \! R_{k l} f^k f^l + ( \ln N )_{| l} \left[ (f^k f_k)^{| l} - f_k (f^{k | l} - f^{l | k} ) - f^l f^k_{| k} \right] + N N^{| k}_{| k}
\nonumber \\ & & \hspace{-10mm} + \frac{1}{2} \frac{N}{\sqrt{g}} g_{k l} \frac{\partial \pi^{k l}}{\partial \tau} + \frac{1}{2} f^{k | l} \frac{\partial g_{k l}}{\partial \tau} + \left ( g^{m n} f^l - \frac{1}{2} g^{l m} f^n \right ) \left [ \left ( \frac{\partial g_{l m}}{\partial \tau} \right )_{| n} \! \! \! - (\ln N)_{| n} \frac{\partial g_{l m}}{\partial \tau} \right ] .
\end{eqnarray}

\noindent Writing out the equation $\delta S / \delta f^k = 0$,
\begin{equation}                                                           
\hspace{-20mm} 0 = \frac{N}{\sqrt{g}} \frac{\delta S}{\delta f^k} = - f_k{}^{| l}_{| l} + f^l{}_{| l k} + \dots ,
\end{equation}

\noindent we have
\begin{eqnarray}\label{ddf-ddf=}                                           
& & \hspace{-20mm} f_k{}^{| l}_{| l} - f^l{}_{| l k} = - ^3 \! R_{k l} f^l + ( \ln N )^{| l} \left ( f_{k | l} + f_{l | k} -2 g_{kl} f^m_{| m} \right ) \nonumber \\ & & + \left ( g^{m n} \delta^l_k - g^{l m} \delta^n_k \right ) \left [ \left ( \frac{\partial g_{l m}}{\partial \tau} \right )_{| n} \! \! \! - (\ln N)_{| n} \frac{\partial g_{l m}}{\partial \tau} \right ] .
\end{eqnarray}

\noindent For $^3 \! R_{k l} = 0$, $N = const$, $\partial g_{k l} / \partial \tau = 0$, these three equations are not independent. They do not determine the longitudinal part of $f^k$ (which is then defined by substituting into (\ref{dd(ff)=})). For a nonzero right-hand side, a consistency condition should be fulfilled. It should be similar to the current conservation law in electrodynamics (now three-dimensional) $[(f_{k | l} - f_{l | k})^{| l}]^{| k} = 0$ imposed on $(f_{k | l} - f_{l | k})^{| l} = f_k{}^{| l}_{| l} - f^l{}_{| l k} - ^3 \! R_{k l} f^l$. Then this condition reads
\begin{eqnarray}\label{ddlnN=R}                                            
& & \left [ ( \ln N )^{| l} \left ( 2 g_{kl} f^m_{| m} - f_{k | l} - f_{l | k} \right ) \right ]^{| k} = \left \{ - 2 \, ^3 \! R_{k l} f^l +  \phantom{\left [ \left ( \frac{1}{2} \right )_{| k} \right ]} \right. \nonumber \\ & & \left. + \left ( g^{m n} \delta^l_k - g^{l m} \delta^n_k \right ) \left [ \left ( \frac{\partial g_{l m}}{\partial \tau} \right )_{| n} \! \! \! - (\ln N)_{| n} \frac{\partial g_{l m}}{\partial \tau} \right ] \right \}^{| k} .
\end{eqnarray}

\noindent This can be considered as an equation for $N$ with a small right-hand side (to be solved iteratively), and actual $f^k$ for the continuum Painlev\'{e}-Gullstrand metric (\ref{PanCartesian}) give a non-degenerate (invertible) second order differential operator of the Laplace type acting on $\ln N$ on the left-hand side.

Six remaining equations $\delta S / \delta g_{k l} = 0$ can be written as
\begin{eqnarray}\label{3R-g3R/2}                                           
& & ^3 \! R^{k l} - \frac{1}{2} g^{k l} \, {}^3 \! R = - \frac{1}{2} \frac{g^{k l}}{g}  \left ( \frac{1}{2} \pi^m{}_m \pi^n{}_n - \pi^m{}_n \pi^n{}_m \right ) \nonumber \\ & & + \frac{2}{g} \left ( \frac{1}{2} \pi^m{}_m \pi^{k l} - \pi^k{}_m \pi^{m l} \right ) + \frac{1}{N \sqrt{g}} \left [ (\pi^{k l} f^m)_{| m} - f^k{}_{| m} \pi^{m l} - f^l{}_{| m} \pi^{m k} \right ]  \nonumber \\ & & + N^{-1} \left ( N^{| k l} - g^{k l} N^{| m}{}_{| m} \right ) - \frac{1}{N \sqrt{g}} \frac{\partial \pi^{k l}}{\partial \tau} .
\end{eqnarray}

\noindent They take the form of the three-dimensional Einstein equations with a non-trivial right-hand side. For the Painlev\'{e}-Gullstrand metric (\ref{PanCartesian}) on the right-hand side, they give $^3 \! R^{k l} = 0$ self-consistently, that is, a flat metric $g_{k l}$. In particular, the considered $f^k$ lead to the zero bilinear contribution on the right-hand side (the sum of terms of the type $(\partial f)(\partial f)$ and $( f)(\partial^2 f)$, the first two lines of equation (\ref{3R-g3R/2})). Simultaneously, these $f^k$ satisfy (\ref{dd(ff)=}) and (\ref{ddf-ddf=}) with zero right-hand side. In (\ref{dd(ff)=}), these $f^k$ vanish the bilinear term $\partial^2 (f f)$ on the left-hand side of (\ref{dd(ff)=}).

It seems a coincidence that $f^k$ satisfy more than three equations with trivial $N-1, g_{k l}$, but this degeneracy is removed for the target finite-difference form of the action, and the remaining variables become non-trivial, albeit small. The terms of the type of $\partial^2 (f f)$, $(\partial f)(\partial f)$ and $( f)(\partial^2 f)$ are related by the product rule of differentiation. If the derivatives $\partial$ are replaced by the finite differences $\Delta$, such a rule continues to hold, but only in the leading order over the differences. If $f^k$ vanish the left-hand side of (\ref{dd(ff)=}) and (\ref{ddf-ddf=}), in particular, the bilinear $\partial^2 (f f)$, the bilinears $(\partial f)(\partial f)$ and $( f)(\partial^2 f)$ on the right-hand side of (\ref{3R-g3R/2}) will be not zero, but $O(\Delta)$. This will lead to $^3 \! R^{k l}$ and metric $g_{k l}$ differing from $\delta_{k l}$ by $O(\Delta)$.

It is important for transition to the discrete form of the equations that the black hole solution be defined without invoking a priory spherical symmetry. We can take $g_{kl} = \delta_{kl}$, $N = 1$, stationarity (independence on $\tau$) and a $\delta$-function-like nature of the source. So we have the system of (\ref{dd(ff)=}) and (\ref{ddf-ddf=}),
\begin{eqnarray}                                                        
& & \label{dd(ff)=0} \frac{1}{2} \nabla^2 \left (\vf^2 \right ) - \frac{1}{2} \left [\nabla \times \vf \right ] \cdot \left [\nabla \times \vf \right ] = 0, \\ & & \label{ddf-ddf=0} \left [\nabla \times [ \nabla \times \vf ] \right ] = 0 ,
\end{eqnarray}

\noindent at $r > 0$. It follows from (\ref{ddf-ddf=0}) that $\vf$ is purely longitudinal, and we write instead of (\ref{dd(ff)=0})
\begin{equation}                                                           
\vf = \nabla \chi , ~~~ \nabla^2 \left ( \left (\nabla \chi \right )^2 \right ) = 0
\end{equation}

\noindent at $r > 0$. This gives
\begin{equation}\label{f^2cont}                                            
\left (\nabla \chi \right )^2 = \frac{r_g}{r} .
\end{equation}

\noindent The constant $r_g$ is chosen to reproduce the Painlev\'{e}-Gullstrand metric. In the form $r \left (\nabla \chi \right )^2 - r_g = 0$, this looks as the Hamilton-Jacobi equation for a particle with the action $\chi$, mass squared $r_g$ and the three-dimensional positively defined metric $r \delta_{kl}$ (more precisely, with all the coordinates having the timelike signature). In spherical coordinates, this reads
\begin{equation}                                                           
\left (\frac{ \partial \chi }{ \partial r} \right )^2 + \frac{1}{r^2} \left (\frac{ \partial \chi }{ \partial \theta} \right )^2 + \frac{1}{r^2 \sin^2 \theta } \left (\frac{ \partial \chi }{ \partial \phi} \right )^2 = \frac{r_g}{r} .
\end{equation}

\noindent According to the standard approach to the Hamilton-Jacobi equation, we try the additive separation of the variables,
\begin{eqnarray}                                                     
& & \chi (r, \theta , \phi ) = R(r) + \Theta (\theta ) + \Phi ( \phi ) , \\ & & \left [ R^\prime ( r ) \right ]^2 + \frac{1}{r^2} \left [ \Theta^\prime ( \theta ) \right ]^2 + \frac{1}{r^2 \sin^2 \theta } \left [\Phi^\prime ( \phi ) \right ]^2 = \frac{r_g}{r} , \\ & & ( \Phi^\prime )^2 = M^2 , ~~~ ( \Theta^\prime )^2 + \frac{M^2 }{\sin^2 \theta } = L^2 , ~~~ ( R^\prime )^2 + \frac{L^2 }{ r^2 } = \frac{r_g}{r} .
\end{eqnarray}

\noindent Here $L^2$, $M^2$ are positive separation constants. It can be seen that $\Theta$ does not exist for sufficiently small values of $\theta$, unless $M = 0$, and $R$ does not exist for sufficiently small values of $r$, unless $L = 0$. Thus, the solution is automatically spherically symmetrical, and we get the Painlev\'{e}-Gullstrand metric.

Thus, equations (\ref{dd(ff)=}), (\ref{ddf-ddf=}) (in particular, the consistency condition (\ref{ddlnN=R})) and (\ref{3R-g3R/2}) form a suitable system for the analysis of a metric close to the Painlev\'{e}-Gullstrand solution if we require that the spherical symmetry of the latter solution would arise automatically without imposing a priori, but after imposing some other conditions. Of these equations, (\ref{dd(ff)=}) seems to be interesting in that it is a combination of the equations for initial conditions and the dynamical equations (in the Hamiltonian formalism terminology). It is interesting to ask what combination of the components of the Einstein equations is this equation. In the variation of the action $\delta S = \int G_{\lambda \mu} \delta g^{\lambda \mu} \d^4 x$, we can express $\delta g^{\lambda \mu}$ in terms of $\delta g_{k l}$, $\delta f^k$, $\delta N$ and find
\begin{eqnarray}                                                           
& & \frac{1}{N \sqrt{g}} \frac{\delta S}{\delta \, ^3 \! g^{k l}} = G_{k l} , ~~~ \frac{N}{2 \sqrt{g}} \frac{\delta S}{\delta f^k} = G_{0k} - G_{k l} f^l , \nonumber \\ & & \frac{N^2}{2 \sqrt{g}} \frac{\delta S}{\delta N} = G_{0 0} - 2 G_{0 k} f^k + G_{k l} f^k f^l ,
\end{eqnarray}

\noindent and the component (\ref{dS/dg-dS/df-dS/dN}) includes all the spatio-spatial, spatio-temporal and temporal-temporal components of the Einstein equations, but looks the simplest in terms of the 4D Ricci tensor,
\begin{eqnarray}                                                           
& & \frac{N}{\sqrt{g}} \left [ \frac{ g_{k l}}{2} \frac{\delta S}{\delta g_{k l}} - f^k \frac{\delta S}{\delta f^k} - \frac{ N}{4} \frac{\delta S}{\delta N} \right ] \nonumber \\ & & = - \frac{1}{2} N^2 \, ^3 \! g^{k l} G_{k l} + \frac{3}{2} f^k f^l G_{k l} - f^k G_{0 k} - \frac{1}{2} G_{0 0} = - R_{0 0} + f^k f^l R_{k l} .
\end{eqnarray}

\noindent Here $^3 \! g^{k l}$ is the inverse of $g_{k l}$.

A situation in which spherical symmetry is absent is exactly the situation with the Regge equations or, here, finite-difference equations. So we use the above input as a definition of a black hole solution ($g_{kl} = \delta_{kl}$, $N = 1$, independence on $\tau$ and a $\delta$-function-like nature of the source). Now we have a finite-difference form of the equations (\ref{dd(ff)=}), (\ref{ddf-ddf=}) (in particular, the consistency condition (\ref{ddlnN=R})) and (\ref{3R-g3R/2}).

In the continuum framework, $r_g$ was an integral characteristic of the $\delta$-function type distribution of the stress-energy tensor with support at $\bx = 0$. In the discrete framework, at large $r$, the solution should be close to the continuum one with a certain $r_g$. Now the point source should show up in that a certain Regge equation should equate the result of varying the action with respect to the (timelike) length of any edge $A_n A_{n+1}$ located along the world line $\bx = 0$ (Fig.~\ref{triangulation}), not to zero, but to a constant. This equation simply serves to express this constant in terms of $r_g$, while the others (those at $\bx \neq 0$) give a nontrivial info on the metric/length values.

The right-hand sides of the latter equations (more exactly, a finite-difference form of (\ref{dd(ff)=}), (\ref{ddf-ddf=}-\ref{3R-g3R/2}) to which they are reduced) differ from zero in the next-to-leading order over the differences $O(\Delta)$. The left-hand sides (including that of the consistency condition (\ref{ddlnN=R}) for $N$, in particular, in the vicinity of the Painlev\'{e}-Gullstrand type metric) can be resolved with respect to $f^k$, $N$ and $^3 \! R_{k l}$ (and eventually $g_{k l}$). An iterative process can be carried out. In the leading (zeroth) order over the differences, we have a finite-difference form of equations (\ref{dd(ff)=0}), (\ref{ddf-ddf=0}), leading to
\begin{equation}\label{DDff}                                               
f_k = a^{-1} \Delta_k \chi , ~~~ \sum^3_{k=1} \bDelta_k \Delta_k \left ( \vf^2 \right ) = \left \{ \begin{array}{rl} 0 & \mbox{at } \bx \neq 0  \\ C & \mbox{at } \bx = 0 . \end{array} \right.
\end{equation}

\noindent Here $\Delta_k h( x^k ) \equiv h(x^k ) - h(x^k - a )$, and the simplest Hermitian form of the difference Laplacian is written out (in the leading order over the differences $\Delta$, such forms are the same); $C$ is a constant chosen so that $\vf$ at large distances would lead to the Painlev\'{e}-Gullstrand metric for a given $r_g$. With the help of passing to the momentum representation, we get
\begin{equation}\label{vf2=}                                               
\vf^2 ( \bx ) = C \int^{ \pi / a }_{\! \! - \pi / a } \int^{ \pi / a }_{\! \! - \pi / a } \int^{ \pi / a }_{\! \! - \pi / a } \frac{\d^3 \bp}{(2 \pi )^3} \frac{\exp (i \bp \, \bx )}{\sum_k 4 \sin^2 (p_k a / 2 )}.
\end{equation}

\noindent At large $r$, small $p$ are significant, $\sum_k 4 \sin^2 (p_k a / 2 ) \approx a^2 \bp^2$, and to reproduce the Painlev\'{e}-Gullstrand metric we must have $C = 4 \pi a^2 r_g$. There also $f_k \approx \partial_k \chi$, and as discussed above, $\chi$ is a function of only $r$, and the required metric is restored.

The function $\vf^2$ could be averaged over the orientation of the Regge manifold with respect to the center $r=0$ and any given point $\bx$. For $\vf^2$ (\ref{vf2=}), this is equivalent to averaging over the orientation of $\bx$ under the integral sign in (\ref{vf2=}),
\begin{eqnarray}\label{<vf2>=}                                             
& & \langle \vf^2 \rangle = \int^{ \pi / a }_{\! \! - \pi / a } \int^{ \pi / a }_{\! \! - \pi / a } \int^{ \pi / a }_{\! \! - \pi / a } \frac{\d^3 \bp}{(2 \pi )^3} \frac{4 \pi a^2 r_g}{\sum_k 4 \sin^2 (p_k a / 2 )} \int \exp (i \bp \, \bx ) \frac{\d^2 \bn }{4 \pi } \nonumber \\ & & = \int^{ \pi / a }_{\! \! - \pi / a } \int^{ \pi / a }_{\! \! - \pi / a } \int^{ \pi / a }_{\! \! - \pi / a } \frac{\d^3 \bp}{(2 \pi )^3} \frac{4 \pi a^2 r_g}{\sum_k 4 \sin^2 (p_k a / 2 )} \frac{\sin pr}{pr} ,
\end{eqnarray}

\noindent where $\bn = \bx / r$. This is the simplest analogue of averaging over the simplicial structures, although ideally we assume that such averaging should apply to physical observables obtained as functions of the metric, should go over various 4-dimensional structures, and not just 3-dimensional leaves, and lead to functions of invariants.

It is also interesting to see this solution in the center $r = 0$. At $\bx = 0$, the formula (\ref{vf2=}) reduces to some table integral \cite{MagOber},
\begin{equation}\label{vf2(0)}                                             
\vf^2 ({\bf 0} ) = \frac{8 r_g}{\pi a} (18 + 12 \sqrt{2} - 10 \sqrt{3} - 7 \sqrt{6} ) K^2 [(2 - \sqrt{3})( \sqrt{3} - \sqrt{2})] \approx 1.05 \frac{\pi r_g}{a} ,
\end{equation}

\noindent $K(k)$ is the complete elliptic integral of the first kind. It is also easy to see that
\begin{eqnarray}\label{vf2(a)}                                             
& & \vf^2 (\pm a, 0, 0) = \vf^2 ( 0, \pm a, 0) = \vf^2 ( 0, 0, \pm a ) = \vf^2 ({\bf 0} ) \nonumber \\ & & - \frac{ 4 \pi a^2 r_g}{6} \int^{ \pi / a }_{\! \! - \pi / a } \int^{ \pi / a }_{\! \! - \pi / a } \int^{ \pi / a }_{\! \! - \pi / a } \frac{\d^3 \bp}{(2 \pi )^3} \approx \left ( 1.05 - \frac{2}{3} \right ) \pi \frac{r_g}{a} \approx 1.19 \frac{r_g}{a} .
\end{eqnarray}

\noindent To find $\chi$, we should solve the equation
\begin{equation}                                                           
a^{-2} \sum^3_{k=1} \left ( \Delta_k \chi \right )^2 = \int^{ \pi / a }_{\! \! - \pi / a } \int^{ \pi / a }_{\! \! - \pi / a } \int^{ \pi / a }_{\! \! - \pi / a } \frac{\d^3 \bp}{(2 \pi )^3} \frac{\pi a^2 r_g \exp (i \bp \bx ) }{\sum_k \sin^2 (p_k a / 2 )} .
\end{equation}

\noindent Looking at (\ref{vf2(a)}), we can assume the validity of the continuum equation for $\chi$ with an accuracy of better than 20\% at the vertices with $r \geq a$ and determine $\chi ( {\bf 0} ) $, knowing $\vf^2 ({\bf 0} )$,
\begin{eqnarray}\label{chi(x)}                                             
& & \chi ( \bx ) = 2 \sqrt{r_g r} \mbox{ at } r \geq a , ~~~ 3 a^{-2} [ \chi ( {\bf 0} )  - \chi ( - a, 0, 0) ]^2 = \vf^2 ( {\bf 0 } ) \approx 1.05 \pi \frac{r_g}{a} , \nonumber \\ & & \chi ( {\bf 0} ) \approx \left ( 2 - \sqrt{\frac{1.05 \pi}{3}} \right ) \sqrt{r_g a} \approx 0.95 \sqrt{r_g a} .
\end{eqnarray}

\noindent Now checking for $\vf^2 (a, 0, 0)$ with this $\chi ( \bx )$ (since the equation for $\chi$ is not symmetrical with respect to the change $a \to - a$ due to the finite differences),
\begin{eqnarray}\label{f(a00)}                                             
& & a^2 \vf^2 (a, 0, 0) = [ \chi (a, 0, 0) - \chi ( {\bf 0} ) ]^2 + [ \chi (a, 0, 0) - \chi (a, -a, 0) ]^2 + [ \chi (a, 0, 0) \nonumber \\ & & - \chi (a, 0, -a) ]^2 \approx \frac{1.05 \pi }{3} r_g a + 8 (1 - 2^{1/4})^2 r_g a \approx 1.39 r_g a .
\end{eqnarray}

\noindent This is at variance with the exact solution (\ref{vf2(a)}) within the same 20\% (but in the opposite direction compared to the continuum formula (\ref{f^2cont})), which seems to be satisfactory for a rough estimate.

Consider such a crude estimate of the effective (shown by the discretized Riemann tensor (\ref{DM+MM})) curvature value at the center. The Riemann tensor components in the continuum GR at $N = 1$, $g_{kl} = \delta_{k l}$ have the form
\begin{eqnarray}\label{R-f}                                          
& & R_{k l m n} = \frac{1}{4} [ (f_{k, m} + f_{m, k} ) (f_{l, n} + f_{n, l} ) - (f_{k, n} + f_{n, k} ) (f_{l, m} + f_{m, l} ) ] , \\ & & R_{0 k l m} = \frac{1}{2} (f_{m, l} - f_{l, m})_{,k} + f_n R_{nklm} , \\ & & \label{Rieman} R_{0k0l} = -\frac{1}{2} (\vf^2)_{, k l} + \frac{1}{4} (f_{m, k} - f_{k, m}) (f_{m, l} - f_{l, m}) + f_m f_n R_{m k n l} ,
\end{eqnarray}

\noindent or, on the solution $f_k = \partial_k \chi$,
\begin{eqnarray}\label{R-chi}                                              
& & R_{k l m n} = \chi_{, k m} \chi_{, l n} - \chi_{, k n} \chi_{, l m}, ~~~ R_{0 k l m} = f_n R_{nklm} , ~~~ R_{0k0l} = -\frac{1}{2} (\vf^2)_{, k l} \nonumber \\ & & + f_m f_n R_{m k n l} .
\end{eqnarray}

\noindent The transition from the derivatives $\partial$ to the finite differences $\Delta$ is not unique, but this non-uniqueness is at the level of higher orders in $\Delta$; their inclusion makes sense if we also take them into account when passing from the original Regge action to the finite-difference action (\ref{DM+MM}). Here we should provide a consistence with the considered field equations, in which we use the simplest Hermitian form of the difference Laplacian (\ref{DDff}). The field equations are combinations of the components of the Ricci tensor and, therefore, of the Riemann tensor. To get such a Laplacian, the second derivative in $R_{0k0l}$ (\ref{Rieman}) should be analogously substituted by such an operator, $\partial_l \partial_k \to - a^{-2} (\bDelta_l \Delta_k + \Delta_l \bDelta_k ) / 2$.

We can go directly to the piecewise-difference form of equations (\ref{R-chi}) (after the substitution $f_k = a^{-1} \Delta_k \chi$, (\ref{DDff}), in (\ref{R-f}-\ref{Rieman})). The normal order of magnitude of the Riemann tensor components at $r_g \ll a$ is $r_g a^{-3}$; $R_{0klm}$ and the second term in $R_{0k0l}$ (\ref{R-chi}) are of a smaller order and, in particular, the Kretschmann scalar in this order is a sum of only positive terms,
\begin{equation}                                                           
R_{\lambda \mu \nu \rho} R^{\lambda \mu \nu \rho} = (R_{k l m n} )^2 + 4 (R_{0 k 0 l} )^2 .
\end{equation}

\noindent Here $g^{\lambda \mu} = {\rm diag} (-1, 1, 1, 1)$ is taken to raise the indices in the leading order over $r_g a^{-1}$. To estimate this at the center, we need to find a few typical finite differences there,
\begin{eqnarray}\label{DDchi}                                              
& & \hspace{-17mm} \Delta_1 \Delta_1 \chi ({\bf 0}) = \chi ({\bf 0}) - 2 \chi (-a, 0, 0) + \chi (-2a, 0, 0) \nonumber \\ & & \approx [(2 - \sqrt{0.35 \pi }) + 2 \sqrt{2} -4] \sqrt{r_g a} \approx -0.22 \sqrt{r_g a}, \nonumber \\ & & \hspace{-17mm} \Delta_2 \Delta_1 \chi ({\bf 0}) = \chi ({\bf 0}) - \chi (-a, 0, 0) - \chi (0, -a, 0) + \chi (-a, -a, 0) \nonumber \\ & & \approx [(2 - \sqrt{0.35 \pi }) + 2 \cdot 2^{1/4} -4] \sqrt{r_g a} \approx -0.67 \sqrt{r_g a}, \nonumber \\ & & \hspace{-17mm} \bDelta_1 \Delta_1 \vf^2 ({\bf 0}) = 2 \vf^2 ({\bf 0}) - \vf^2 (-a, 0, 0) - \vf^2 (a, 0, 0) = \frac{4}{3} \pi \frac{r_g}{a} \approx 4.19 \frac{r_g}{a}, \nonumber \\ & & \hspace{-17mm} ( \bDelta_2 \Delta_1 + \bDelta_1 \Delta_2 ) \vf^2 ({\bf 0}) = 2 \vf^2 ({\bf 0}) + \vf^2 (-a, a, 0) + \vf^2 (a, -a, 0) \nonumber \\ & & - \vf^2 (-a, 0, 0) - \vf^2 (0, a, 0) - \vf^2 (a, 0, 0) - \vf^2 (0, -a, 0) \nonumber \\ & & \approx 2 \cdot 1.05 \pi \frac{r_g}{a} + 2 \frac{r_g}{a \sqrt{2}} - 4 \left ( 1.05 - \frac{2}{3} \right ) \pi \frac{r_g}{a} \approx 3.19 \frac{r_g}{a} .
\end{eqnarray}

\noindent This yields typical components of the discrete Riemann tensor of interest,
\begin{equation}                                                           
R_{1212} \approx - 0.40 \frac{r_g}{a^3}, ~~~ R_{1213} \approx - 0.30 \frac{r_g}{a^3}, ~~~ R_{0101} \approx 2.09 \frac{r_g}{a^3}, ~~~ R_{0102} \approx 0.80 \frac{r_g}{a^3} ,
\end{equation}

\noindent and the required Kretschmann scalar,
\begin{equation}\label{RR0}                                                
R_{\lambda \mu \nu \rho} R^{\lambda \mu \nu \rho} ({\bf 0}) \! = 12 (R_{1212} )^2 + 24 (R_{1213} )^2 + 12 (R_{0101} )^2 + 24 (R_{0102} )^2 \approx 71.9 \frac{r_g^2}{a^6}.
\end{equation}

\noindent It is to be compared with that one at large distances (that is, the continuum value),
\begin{equation}\label{RRSchw}                                             
R_{\lambda \mu \nu \rho} R^{\lambda \mu \nu \rho} = 12 \frac{r_g^2}{r^6}.
\end{equation}

\noindent (\ref{RR0}) looks like (\ref{RRSchw}), cut off at $r$ slightly less than $a$ ($r = 6^{- 1 / 6} a = 0.74 a$).

The Schwarzschild metric follows approximately at large distances, $r^3 \gg a^2 r_g$, when $\vf \approx \bx (r_g / r^3)^{1 / 2}$, by the standard additive redefinition of $\tau$ by a certain function of $r$.

If $r_g r^{-3} a^2 \sim 1$, then the defect angles are also $\sim 1$, and probably we need to directly solve the Regge skeleton equations instead, and the above consideration reformulated in terms of edge lengths can give some initial approximation for this.

\section{Conclusion}

Remarkable is that the Schwarzschild geometry can be obtained from general Einstein equations without requiring a priori spherical symmetry. This allows to formulate the discrete version of the Schwarzschild problem and get its solution which tends at sufficiently large distances to the Painlev\'{e}-Gullstrand metric, which is described by pure shift ADM functions. The equations obtained from the 3 + 1 ADM formalism are convenient for this, this time not in the Hamiltonian form, but in the Lagrangian form. It is appropriate to solve iteratively these equations rewritten in a certain form (the black hole solution follows just in the zeroth order). Namely, the dynamical equations (for $\partial g_{k l} / \partial \tau$) turn into 3D Einstein equations for $g_{k l}$ with certain right-hand sides; the equations for initial conditions, Hamiltonian and diffeomorphism constraints (of which the Hamiltonian constraint being combined in a certain way with the dynamical equations and the diffeomorphism constraints) turn into the equations for the lapse-shift functions (here $N, f^k$). This scheme also seems convenient for analyzing any small perturbations of the Schwarzschild geometry in the Painlev\'{e}-Gullstrand coordinates, and not just its discretization.

Passing to the simplest periodic Regge lattice consists in rewriting the GR action in the finite-difference form in the leading order over metric variations from simplex to simplex. A complication is that fixing the edge lengths implies loose dynamical fixation of the spacelike lengths while the discrete lapse-shifts should be fixed manually, like gauge conditions. Therefore, the metric (in the 3+1 ADM form) should be transformed to the synchronous frame (like the Lemaitre frame in the present context) and only then substituted into the finite-difference form of the action. Moreover, in the particular case of the leading order over metric variations or over finite differences, the action in the transformed coordinates reduces to the action in the original coordinates with the 3+1 ADM metric by invariance, and we return to the above ADM equations, now in the discrete form and with the dynamically fixed elementary length scale, and get the discrete Painlev\'{e}-Gullstrand metric with the resolved singularity.

Turning to a comparison with the result of the resolution of the Schwarzschild black hole singularity in Loop Quantum Gravity \cite{Ash1,Ash2}, we note that the quantum of area or area gap responsible for this resolution is proportional to the Barbero-Immirzi parameter $\gamma$. In our case, we have a discrete version of the Barbero-Immirzi parameter $\gamma$, and the factor in the functional measure $F$ at $\gamma \ll 1$ has a local maximum at the elementary area scale $a^2$ proportional to $\gamma$ (and to Planck scale, of course). However, there is another maximum, which at the value of another parameter $\eta \gg 1$ arises at $a^2$ proportional to $\eta$ (\ref{a=sqrt}), and the maximum at $a^2 \sim \gamma$ is negligibly smaller \cite{our1}. Then the resolution of the singularity in our case occurs on a certain scale $a$ larger than the Planck scale, at least formally.

We have considered the point matter source. A description of the electromagnetic field on a simplicial manifold was developed in \cite{Sor,Wein}, which allows for the discrete analysis with this example of an extended matter source.

\section*{Acknowledgments}

The present work was supported by the Ministry of Education and Science of the Russian Federation.


\begin{thebibliography}{99}
\bibitem{Regge}
 T. Regge, General relativity theory without coordinates, {\it Nuovo Cimento} {\bf 19}, 558 (1961).
\bibitem{Fein}
 G. Feinberg, R. Friedberg, T. D. Lee, and M. C. Ren, Lattice gravity near the continuum
 limit, {\it Nucl. Phys. B} {\bf 245}, 343 (1984).
\bibitem{CMS}
 J. Cheeger, W. M\"{u}ller, and R. Shrader, On the curvature of the piecewise flat spaces,
 {\it Commun. Math. Phys.} {\bf 92}, 405 (1984).
\bibitem{WilTuc}
 R. M. Williams and P. A. Tuckey, Regge calculus:  a brief review and bibliography, {\it Class. Quantum Grav.} {\bf 9}, 1409 (1992).
\bibitem{RegWil}
 T. Regge and R. M. Williams, Discrete structures in gravity, {\it Journ. Math. Phys.} {\bf 41}, 3964 (2000); ({\it Preprint} arXiv:gr-qc/0012035).
\bibitem{Ham}
 H. W. Hamber, Quantum Gravity on the Lattice, {\it Gen. Rel. Grav.} {\bf 41}, 817 (2009); ({\it Preprint} arXiv:0901.0964[gr-qc]).
\bibitem{cdt}
 J. Ambjorn, A. Goerlich, J. Jurkiewicz, and R. Loll,  Nonperturbative Quantum Gravity, {\it Physics Reports} {\bf 519}, 127 (2012); ({\it Preprint} arXiv:1203.3591[hep-th]).
\bibitem{Per}
 A. Perez, The spin-foam approach to quantum gravity, {\it Living Rev. Relativity} {\bf 16} (2013), DOI: 10.12942/lrr-2013-3; ({\it Preprint} arXiv:1205.2019[gr-qc]).
\bibitem{Dup}
 M. Dupuis, J. P. Ryan, and S. Speziale, Discrete gravity models and Loop Quantum Gravity: a short review, {\it SIGMA} {\bf 8}, 052 (2012); ({\it Preprint} arXiv:1204.5394[gr-qc]).
\bibitem{Wong}
 C.-Y. Wong, Application of Regge calculus to the Schwarzshild and Reissner-Nordstrøm geometries, {\it Journ. Math. Phys.} {\bf 12}, 70 (1971).
\bibitem{Bre2}
 L. Brewin, Einstein-Bianchi system for smooth lattice general relativity. I. The Schwarzschild spacetime, {\it Phys. Rev. D} {\bf 85}, 124045 (2012); ({\it Preprint} arXiv: $\! \! \! \! $ 1101.3171[gr-qc]).
\bibitem{WilCol}
 P. A. Collins and R. M. Williams, Dynamics of the Friedmann universe using Regge calculus,  {\it Phys. Rev. D} {\bf 7}, 965 (1973).
\bibitem{Gen}
 A. P. Gentle, A cosmological solution of Regge calculus, {\it Class. Quantum Grav.} {\bf 30}, 085004 (2013); ({\it Preprint} arXiv:1208.1502[gr-qc]).
\bibitem{Bre4}
 L C Brewin, A numerical study of the Regge calculus and Smooth Lattice methods on a Kasner cosmology, {\it Classical and Quantum Gravity} {\bf 32}, 195008 (2015); ({\it Preprint} arXiv:1505.00067[gr-qc]).
\bibitem{WilLiu}
 R. G. Liu and R. M. Williams, Regge calculus models of closed lattice universes, {\it Phys. Rev. D} {\bf 93}, 023502 (2016); ({\it Preprint} arXiv:1502.03000[gr-qc]).
\bibitem{GlaLol}
  L. Glaser and R. Loll, CDT and cosmology, {\it Comptes Rendus Physique} {\bf 18}, 265 (2017); {\it Preprint} arXiv:1703.08160[gr-qc])
\bibitem{Ash1}
 A. Ashtekar, J. Olmedo and P. Singh, Quantum  Transfiguration  of Kruskal Black Holes, {\it Phys. Rev. Lett.} {\bf 121}, 241301 (2018); {\it Preprint} arXiv:1806.00648[gr-qc]).
\bibitem{Ash2}
 A. Ashtekar, J. Olmedo and P. Singh, Quantum  extension  of  the Kruskal spacetime, {\it Phys. Rev. D} {\bf 98}, 126003 (2018); {\it Preprint} arXiv:1806.02406[gr-qc]).
\bibitem{our1}
 V. M. Khatsymovsky, On the non-perturbative graviton propagator, {\it Int. J. Mod. Phys. A} {\bf 33}, 1850220 (2018); ({\it Preprint} arXiv:1804.11212).
\bibitem{Fro}
 J. Fr\"{o}hlich, Regge calculus and discretized gravitational functional integrals, in {\it Nonperturbative Quantum Field Theory: Mathematical Aspects and Applications, Selected Papers} (World Scientific, Singapore, 1992), p. 523, IHES preprint 1981 (unpublished).
\bibitem{ADM}
 R. Arnowitt, S. Deser, and C. W. Misner, Canonical variables for general relativity, {\it Phys. Rev.} {\bf 117}, 1595 (1960).
\bibitem{ADM1}
 R. Arnowitt, S. Deser, and C. W. Misner, The Dynamics of General Relativity, in {\it Gravitation: an introduction to current research, Louis Witten ed.} (Wiley, 1962), chapter 7, p. 227; ({\it Preprint} arXiv:gr-qc/0405109).
\bibitem{Mis}
 C. W. Misner, Feynman quantization of general relativity, {\it Rev. Mod. Phys.} {\bf 29}, 497 (1957).
\bibitem{DeW}
 B. S. DeWitt, Quantization of fields with infinite-dimensional invariance groups. III.
 Generalized Shwinger-Feynman theory, {\it Journ. Math. Phys.} {\bf 3}, 1073 (1962).
\bibitem{our2}
 V. M. Khatsymovsky, On the discrete Christoffel symbols, {\it Int. J. Mod. Phys. A} {\bf 34}, 1950186 (2019); ({\it Preprint} arXiv:1906.11805).
\bibitem{Lemaitre}
 G. Lemaitre, {\it Ann. Soc. Sci. Bruxelles} A {\bf 53}, 51 (1933).
\bibitem{Stan}
 K. P. Stanyukovich, On the question of the Schwarzschild metric in a synchronous reference frame, {\it Reports of the USSR Acad. Sci.} {\bf 187}, 75 (1969).
\bibitem{Painleve}
 P. Painlev\'{e}, {\it C. R. Acad. Sci. (Paris)} {\bf 173} (October 24), 677 (1921).
\bibitem{Gullstrand}
 A. Gullstrand, {\it Arkiv. Mat. Astron. Fys.} {\bf 16(8)}, 1 (1922).
\bibitem{MisTorWhe}
 C. W. Misner, K. S. Thorne, and J. A. Wheeler, {\it Gravitation} (Freeman and Company, San Francisco, 1973).
\bibitem{MagOber}
 W. Magnus und F. Oberhettinger, {\it Formeln und Satze fur die speziellen Funktionen der mathematischen Physik} (Springer Verlag Berlin - G\"{o}ttingen - Heidelberg, 1948).
\bibitem{Sor}
 R. Sorkin, The electromagnetic field on a simplicial net, {\it Journ. Math. Phys.} {\bf 16}, 2432 (1975).
\bibitem{Wein}
 Don Weingarten, Geometric formulation of electrodynamics and general relativity in discrete space-time, {\it Journ. Math. Phys.} {\bf 18}, 165 (1977).
\end{thebibliography}
\end{document}